\documentclass[letters,fleqn,usenatbib]{mnras}

\usepackage{newtxtext,newtxmath}

\usepackage[T1]{fontenc}

\DeclareRobustCommand{\VAN}[3]{#2}
\let\VANthebibliography\thebibliography
\def\thebibliography{\DeclareRobustCommand{\VAN}[3]{##3}\VANthebibliography}

\usepackage{graphicx}	
\usepackage{amsmath}	
\usepackage{array,multirow,graphicx}
\usepackage{relsize}
\usepackage{makecell}
\usepackage{xcolor}
\usepackage[utf8]{inputenc}

\title[Phaethon polarimetry in the negative branch]{(3200) Phaethon Polarimetry in the Negative Branch: New Evidence for the Anhydrous Nature of the DESTINY$^+$ Target Asteroid}

\author[J. Geem et al.]{Jooyeon Geem,$^{1,2} $ \thanks{E-mail: ksky0422@snu.ac.kr}
Masateru Ishiguro,$^{1,2} $ \thanks{E-mail: ishiguro@snu.ac.kr}
Jun Takahashi,$^{3}$
Hiroshi Akitaya,$^{4} $
Koji S. Kawabata,$^{5}$
\newauthor
Tatsuya Nakaoka,$^{5}$
Ryo Imazawa,$^{6}$
Fumiki Mori,$^{6}$
Sunho Jin,$^{1,2}$
Yoonsoo P. Bach,$^{1,2}$
Hangbin Jo,$^{1,2}$
\newauthor
Daisuke Kuroda,$^{7} $
Sunao Hasegawa,$^{8}$
Fumi Yoshida,$^{4,9} $
Ko Ishibashi,$^{4} $
Tomohiko Sekiguchi,$^{10} $
\newauthor
Jin Beniyama, $^{11,12}$
Tomoko Arai,$^{4} $
Yuji Ikeda,$^{13,14} $
Yoshiharu Shinnaka,$^{13} $
Mikael Granvik,$^{15,16}$
\newauthor
Lauri Siltala,$^{15}$
Anlaug A. Djupvik,$^{17,18}$
Anni Kasikov,$^{17,18,19}$
Viktoria Pinter,$^{17,18,20}$
Emil Knudstrup$^{17,18}$ 
\\
\\
$^{1} $Department of Physics and Astronomy, Seoul National University, 1 Gwanak-ro, Gwanak-gu, Seoul 08826, Republic of Korea\\
$^{2} $SNU Astronomy Research Center, Seoul National University, 1 Gwanak-ro, Gwanak-gu, Seoul 08826, Republic of Korea\\
$^{3} $Center for Astronomy, University of Hyogo, 407-2 Nishigaichi, Sayo, Hyogo 679-5313, Japan\\
$^{4} $Planetary Exploration Research Center, Chiba Institute of Technology, 2-17-1 Tsudanuma, Narashino, Chiba 275-0016, Japan\\
$^{5} $Hiroshima Astrophysical Science Center, Hiroshima University, Kagamiyama 1-3-1, Higashi-Hiroshima, Hiroshima 739-8526, Japan\\
$^{6} $Department of Physics, Graduate School of Advanced Science and Engineering, Hiroshima University, Kagamiyama, 1-3-1, Higashi-Hiroshima, Hiroshima\\
$^{} $ 739-8526, Japan\\
$^{7} $Okayama Observatory, Kyoto University, 3037-5 Honjo, Kamogata, Asakuchi, Okayama 719-0232, Japan\\
$^{8} $Institute of Space and Astronautical Science (ISAS), Japan Aerospace Exploration Agency (JAXA), Sagamihara, Kanagawa 252-5210, Japan\\
$^{9} $School of Medicine, University of Occupational and Environmental Health, 1-1 Iseigaoka, Yahata, Kitakyusyu 807-8555, Japan\\
$^{10} $Asahikawa Campus, Hokkaido University of Education, Hokumon, Asahikawa, Hokkaido 070-8621, Japan\\
$^{11}$Institute of Astronomy, Graduate School of Science, The University of Tokyo, 2-21-1 Osawa, Mitaka, Tokyo 181-0015, Japan\\
$^{12}$Department of Astronomy, Graduate School of Science, The University of Tokyo, 7-3-1 Hongo, Bunkyo-ku, Tokyo 113-0033, Japan\\
$^{13}$Koyama Astronomical Observatory, Kyoto Sangyo University, Motoyama, Kamigamo, Kita-Ku, Kyoto, Kyoto 603-8555, Japan\\
$^{14} $Photocoding, 460-102 Iwakura-Nakamachi, Sakyo-ku, Kyoto, 606-0025, Japan\\
$^{15}$Department of Physics, University of Helsinki, PO. Box 64, FI-00014 Helsinki, Finland\\
$^{16}$Asteroid Engineering Laboratory, Lule\aa\, University of Technology, Box 848, SE-98128 Kiruna, Sweden\\
$^{17}$Nordic Optical Telescope, Rambla Jos\'{e} Ana Fern\'{a}ndez P\'{e}rez 7, ES-38711 Bre\~{n}a Baja, Spain\\
$^{18}$Department of Physics and Astronomy, Aarhus University, Ny Munkegade 120, DK-8000 Aarhus C, Denmark\\
$^{19}$Tartu Observatory, University of Tartu, Observatooriumi 1, T\~{o}ravere, 61602, Estonia\\
$^{20}$Physics Department, University of Craiova, Alexandru Ioan Cuza 13, 200585 Craiova, Romania}

\date{Accepted 2022 June 30. Received 2022 June 17; in original form 2022 April 21}

\pubyear{2022}

\begin{document}
\label{firstpage}
\pagerange{\pageref{firstpage}--\pageref{lastpage}}
\maketitle

\begin{abstract}
We report on the first polarimetric study of (3200) Phaethon, the target of JAXA's {\it DESTINY$^+$} mission, in the negative branch to ensure its anhydrous nature and to derive an accurate geometric albedo. We conducted observations at low phase angles (Sun-target-observer angle, $\alpha=8.8$--$32.4\degr$) from 2021 October to 2022 January and found that Phaethon has a minimum polarization degree $P_\mathrm{min}=-1.3 \pm 0.1$\,\%, a polarimetric slope $h= 0.22 \pm 0.02$\,\%\,$\mathrm{deg}^{-1}$, and an inversion angle $\alpha_0= 19.9 \pm 0.3\degr$. The derived geometric albedo is $p_{V} = 0.11$ (in the range of $0.08$--$0.13$). These polarimetric properties are consistent with anhydrous chondrites, and contradict hydrous chondrites and typical cometary nuclei.

\end{abstract}



\section{Introduction}

C-complex asteroids are particularly important for revealing the aqueous activity that might have occurred $<$ 10\, Myr after the beginning of the solar system formation \citep{2012NatCo...3..627F}. Most of them are rich in volatile components, maintaining the primordial information since the formation epoch \citep{2015aste.book..635K}. Accordingly, recent asteroid explorations targeted carbonaceous asteroids. The {\it OSIRIS-REx} mission investigated its target asteroid (101955) Bennu and revealed unambiguous evidence for widespread hydrated minerals \citep{Hamilton+2019}. On the other hand, (162173) Ryugu, the target asteroid of the {\it Hayabusa2} mission, indicated a weak signature of the hydrated minerals that might have experienced a mild heating process at $>$ 300\,$^\circ$C in the parent body \citep{Kitazato+2021}. Therefore, hydrated silicate abundance is an important tracer for the thermal history of C-complex asteroids \citep{Hiroi+1996}.

(3200) Phaethon (F- or B-type, a subclass of C-complex, \citealt{Tholen+1989,Bus+2002}) is the target of JAXA's {\it DESTINY$^+$} mission \citep{Arai+2018}, and known to have unique properties. It has an asteroid-like orbit (the Tisserand parameter with respect to Jupiter, $T_\mathrm{J}>3$) that likely originates in the main asteroid belt \citep{deLeon+2010,MacLennan+2021}. It has shown evidence for dust ejection reminiscent of comets \citep{Jewitt+2010}. Phaethon's albedo has not been determined well, making it difficult to identify if this object consists of a comet-like or asteroid-like composition (see Section \ref{sec:results}).

There is a large discrepancy in the interpretation of Phaethon's spectrum. \citet{Licandro+2007} argued that Phaethon's spectrum is similar to those of aqueously altered CI/CM meteorites and hydrated minerals. \citet{Licandro+2007} further suggested that Phaethon is likely an activated asteroid similar to the main-belt comets rather than typical comets of outer solar system origins. On the other hand, \citet{Clark+2010} reported that Phaethon's spectrum matches CK meteorites or an experimental mixture of chlorite and carbon lampblack. Later, \citet{Takir+2020} reported that this asteroid shows no hydrated mineral absorption near 3\,$\mu$m, supporting the idea of anhydrous material. Note that the interpretation of anhydrous material conflicts with \citet{Licandro+2007}. Such a large discrepancy arises the necessity to examine the nature of Phaethon by a method independent of spectroscopy.

Recently, \citet{Ishiguro2022} proposed that polarimetry at low phase angles (Sun--target--observer angle, $\alpha \lesssim 20 \degr$) is a useful diagnostic tool for conjecturing if C-complex asteroids are hydrous or anhydrous. However, due to the unfavorable observational conditions until recently, Phaethon's polarimetric property at low phase angles ($\alpha < 19.1 \degr$) has not been investigated. Taking advantage of the opportunity in late 2021 and early 2022, we obtained polarimetry at low phase angles ($\alpha= 8.8$--$32.4\degr$) and found that Phaethon's surface is anhydrous. In addition, we narrowed down the albedo estimate range with our polarimetry. 
In this paper, we describe our observations in Section \ref{sec:observations} and the derivation of polarimetric parameters in Section \ref{sec:polpara}. We provide two major findings (the composition and geometric albedo) in Section \ref{sec:results}. We discuss these results in Section \ref{sec:discussion}, focusing on the significance of the albedo determination and hydrous/anhydrous nature.

\section{Observations and data analysis}
\label{sec:observations}

We made polarimetric observations using three instruments: the Hiroshima Optical and Near-InfraRed camera (HONIR; \citealt{Akitaya+2014}) on the 1.5-m Kanata Telescope at the Higashi-Hiroshima Observatory, the Wide Field Grism Spectrograph 2 (WFGS2; \citealt{Uehara+2004,Kawakami+2021}) on the 2.0-m Nayuta telescope at the Nishi-Harima Astronomical Observatory, and the Andalucia Faint Object Spectrograph and Camera (ALFOSC) with the FAPOL polarimeter on the 2.56-m Nordic Optical Telescope at the Observatorio del Roque de los Muchachos, La Palma. These instruments equip a polarizer and a rotatable half-wave plate mounted in the Cassegrain focus of each telescope. We acquired HONIR and WFGS2 data at four different angles of the half-wave plate (0\degr, 45\degr, 22.5\degr, and 67.5\degr, in that order) and FAPOL data at 16 different angles (0\degr, 22.5\degr, 45\degr, 67.5\degr, 90\degr, 112.5\degr, 135\degr, 157.5\degr, 180\degr, 202.5\degr, 225\degr, 247.5\degr, 270\degr, 292.5\degr, 315\degr, and 337.5\degr, in that order). We used only $R_\mathrm{C}$-band filter. In addition to these new observations, we reanalyzed the $R_\mathrm{C}$-band polarimetric data published in \citet{Shinnaka+2018}. Because this data was taken near the inversion angle with a good signal-to-noise (S/N) ratio (i.e., small random errors), we reanalyzed them by paying particular attention to the systematic errors.

An outline of the data analysis consists of five major steps: (1) preprocessing the raw observed images, (2) extraction of source signals by aperture photometry, (3) correction for systematic errors, (4) derivation of Stokes parameters ($q$ and $u$), polarization degree ($P$), and polarization position angle ($\theta_\mathrm{P}$), and (5) obtaining the nightly weighted mean of $q$ and $u$. Because we strictly followed the reduction processes (1), (2), (4), and (5) written in \citet{Ishiguro2022}, we skipped the detailed explanation in this paper. The reduction process (3) is particularly important for this work, not only because the polarization degrees at these phase angles are small ($P \lesssim 1-2$ \%) and comparable to the instrumental polarization (an inherent artifact of polarization) of some instruments but also because we need to compare the data taken with different instruments. 

In the HONIR data analysis, we examined the polarization efficiency ($P_\mathrm{eff}$) by observing a star (HD 14069) through a wire-grid filter. We investigated the instrumental polarization parameters ($q_\mathrm{inst}$ and $u_\mathrm{inst}$) and position angle offset ($\theta_\mathrm{off}$) through observations of unpolarized stars (G191B2B, HD 212311, and BD +32 3739) and strongly polarized stars (HD 29333, BD +59 389, BD +64 106, and HD 204827). We determined $P_\mathrm{eff}$ = $97.58 \pm 0.08$ \%, $q_\mathrm{inst} = -0.0097 \pm 0.0498$ \%, $u_\mathrm{inst} = -0.0077 \pm 0.0371$ \%, and $\theta_\mathrm{off} = 36.08 \pm 0.13\degr$. These parameters are consistent with \citet{Akitaya+2014}, ensuring the long-term stability of the polarimetric performance of HONIR.
 
In the WFGS2 data analysis, it was reported that $q_\mathrm{inst}$ and $u_\mathrm{inst}$ depended on the instrument rotator angle ($\theta_\mathrm{rot}$). To eliminate this effect, we observed unpolarized stars (HD 212311 and HD 21447) at four different instrument rotator angles and derived two equations: $q_\mathrm{inst}(\theta_\mathrm{rot}) = q_{0}\cos{2\theta_\mathrm{rot}} - u_{0}\sin{2\theta_\mathrm{rot}}$ and $u_\mathrm{inst}(\theta_\mathrm{rot}) = q_{0}\sin{2\theta_\mathrm{rot}} + u_{0}\cos{2\theta_\mathrm{rot}}$, where $q_\mathrm{0} = -0.042 \pm 0.016$ \% and $u_{0} = 0.178 \pm 0.011$ \% for the 2021 October observation and $q_{0} = -0.043 \pm 0.012$ \% and $u_{0} = 0.273 \pm 0.012$ \% for the 2021 November observation. We determined $\theta_\mathrm{off} = -5.19 \pm 0.15\degr$ from the observations of strongly polarized stars (HD 204827, HD 25443, BD +59 389, and HD 19820). We assumed $P_\mathrm{eff}$ $=1$.

In the FAPOL data analysis, we divided each set of data (consisting of 16 different half-wave plate angles data) into four subgroups. The procedure for deriving the Stokes parameters from each subgroup (the process (4)) is the same as the procedure for HONIR and WFGS2. To investigate $q_\mathrm{inst}$, $u_\mathrm{inst}$, and $\theta_\mathrm{off}$, two unpolarized stars (G191B2B and HD 14069) and one strongly polarized star (BD +59 389) were observed. We determined $q_\mathrm{inst} = -0.05 \pm 0.07$ \%, $u_\mathrm{inst} = -0.04 \pm 0.11$ \%, and $\theta_\mathrm{off}=-92.30 \pm 0.06\degr$. These values are in good agreement with previous observations \citep{Ishiguro2022}.

We analyzed the PICO data following \citet{Ikeda+2007}. However, it should be noted that $q_\mathrm{inst}$ and $u_\mathrm{inst}$ errors in our analysis are different from those described in \citet{Ikeda+2007}. They estimated the errors of the instrumental polarization to be $\sim 0.3\, \% $ over the entire field of view (5\arcmin$\times$10\arcmin). After analyzing standard star data taken during the Phaethon observations, we found that the instrumental polarization of PICO was significantly smaller than $0.3\, \%$. The Phaethon's images were taken in the central part of PICO, where the polarization performance is the best in the field of view \citep{Ikeda+2007}. Accordingly, we considered 0.1 \% errors for $q_\mathrm{inst}$ and $u_\mathrm{inst}$ and derived Phaethon's polarization degrees. We also updated the errors of $P_\mathrm{eff}$ and $\theta_\mathrm{off}$ to 0.02\,\% and 0.18$\degr$ based on the measurement of calibration data taken during Phaethon's run. Although we only use data at a low phase angle ($\alpha < 30^\circ$), we confirm that our results show good agreement with \citet{Shinnaka+2018} within their $3\sigma$-uncertainty throughout the whole phase angles. Only the errors are slightly different because we considered systematic errors comprehensively, following the data reduction processes in \citet{Ikeda+2007}.

\begin{figure}
	\includegraphics[width=\columnwidth]{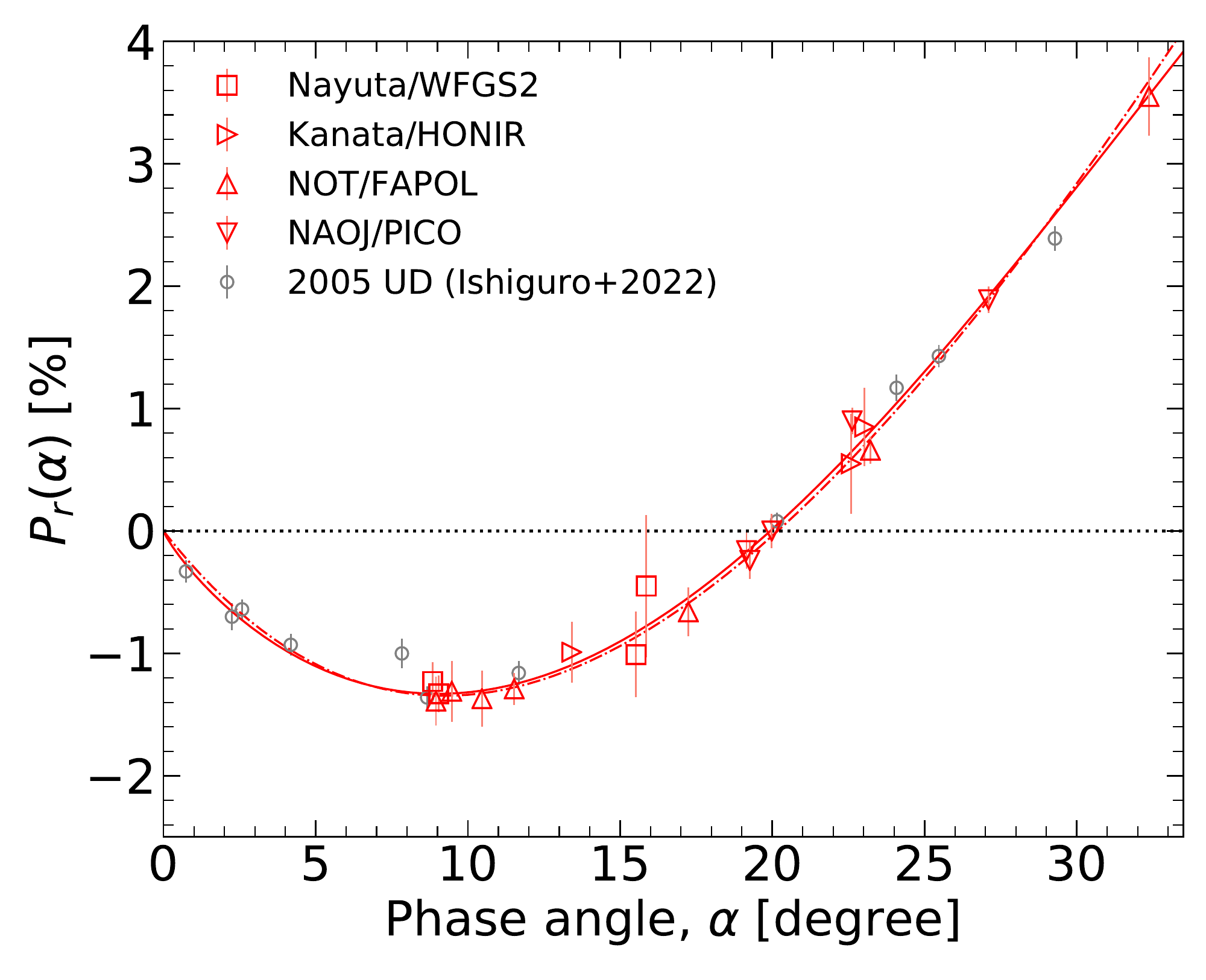}
    \caption{Phase angle ($\alpha$) dependence of polarization degree ($P_\mathrm{r}$). The solid and dash-dot lines are curves that fit only the Phaethon data using the trigonometric and linear-exponential functions, respectively.
    }
    \label{fig:phase-plot}
\end{figure}

\begin{figure}
	\includegraphics[width=\columnwidth]{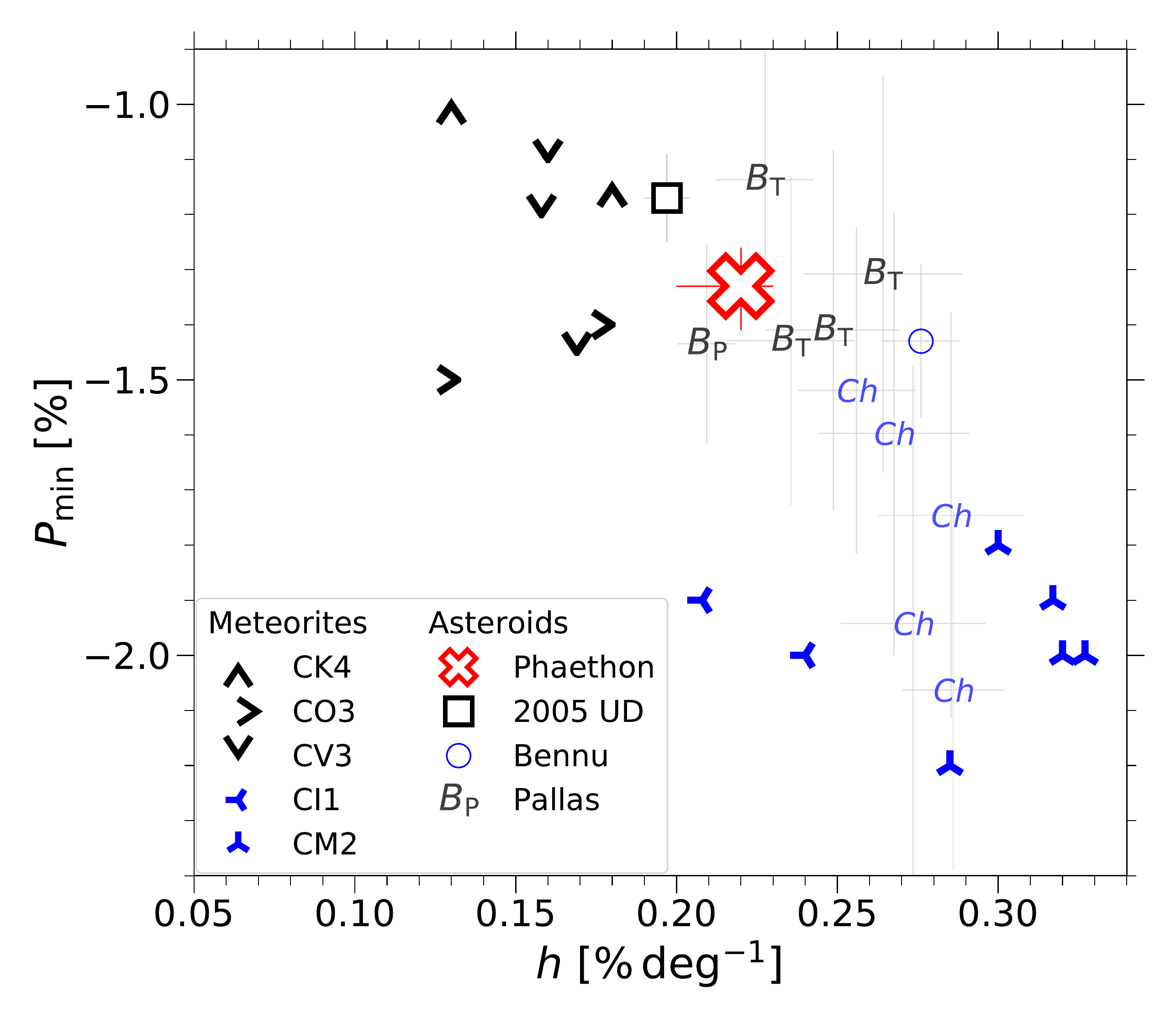}
    \caption
    {Comparison of $h$--$P_\mathrm{min}$ between Phaethon, other B-type asteroids, and carbonaceous meteorites. $B_\mathrm{T}$ and $B_\mathrm{P}$ indicate B-type asteroids with a low albedo (i.e., the Themis group) and a high albedo (i.e., (2) Pallas belongs to the Pallas group, \citealt{Clark+2010}). Ch-type is based on the classification in \citet{Bus+2002}. The meteorites and asteroid data are obtained from \citet{Zellner+1977}, \citet{Geake+1986},\citet{Cellino2018} and \citet{Gil-Hutton+2014}. F-type asteroids are not shown because of large $h$ values.}
    \label{fig:h-Pmin}
\end{figure}

\begin{table*}
\caption{Observation Circumstance and Polarimetric Result}
\label{table:obs&result}
\begin{tabular}{lccccccccccccc}
\hline
Date in UT$^a$  & Inst$^b$ & Exp$^c$ &N$^d$ &  $ r^e $ & $ \Delta^f $ &  $ \phi^g $& $ \alpha^h$&$P^i$&$\sigma \,P^j$&$\theta_{P}^k$&$\sigma \, \theta_{P}^l$ &${P_\mathrm{r}}^m$&${\theta_\mathrm{r}}^n$ \\
 &  & (s)& & ($ \mathrm{au} $) & ($ \mathrm{au} $) & ($\degr$) & ($\degr$)&($\%$)&($\%$)&($\degr$)&($\degr$)&($\%$)&($\degr$)\\
   \hline

2021 Oct 27 18:53--19:35 & WFGS2 &  300 &8   &  2.31  &  1.46 &  235.4 &  15.9  & 0.51 & 0.58 & 69.9 & 47.2 & -0.45 & 104.5 \\
2021 Oct 28 14:41--18:14 & WFGS2 &  300 &12  &  2.31  &  1.45 &  234.1 &  15.5  & 1.09 & 0.35 & 65.3 & 16.3 & -1.01 & 101.2 \\
2021 Nov 14 12:35--20:31 & WFGS2 &  300 &40  &  2.25  &  1.30 &  188.3 &   9.0  & 1.47 & 0.15 & 21.1 & 10.8 & -1.33 & 102.8 \\
2021 Nov 15 10:24--15:23 & WFGS2 &  300 &44  &  2.25  &  1.30 &  184.0 &   8.8  & 1.29 & 0.16 & 12.4 & 14.3 & -1.23 &  98.4 \\
2021 Nov 02 17:22--19:50 & HONIR &  120 &56  &  2.29  &  1.39 &  225.3 &  13.4  & 1.00 & 0.25 & 41.0 &  7.3 & -0.99 &  85.7 \\
2021 Dec 22 10:22--13:17 & HONIR &  120 &64  &  2.07  &  1.33 &   80.5 &  22.6  & 0.57 & 0.41 &-0.29 & 20.3 &  0.55 &   9.2 \\
2021 Dec 23 09:12--12:41 & HONIR &  120 &80  &  2.06  &  1.34 &   79.8 &  23.0  & 0.96 & 0.32 &  3.6 &  9.6 &  0.85 &  13.8 \\
2021 Nov 10 00:57--02:49 & FAPOL & 180 & 16 & 2.27   & 1.33  & 206.2  & 10.5   & 1.37 & 0.23 & 26.5 & 14.9 & -1.37 &  90.4 \\
2021 Nov 13 01:18--01:33 & FAPOL & 180 & 8  & 2.26   & 1.31  & 195.2  & 9.5    & 1.33 & 0.25 & 10.7 &  7.9 & -1.31 &  85.5 \\
2021 Nov 15 01:55--02:23 & FAPOL & 180 & 8  & 2.25   & 1.30  & 186.4  & 9.0    & 1.47 & 0.20 & 176.7&  5.5 & -1.39 &  80.3 \\
2021 Nov 30 22:15--23:33 & FAPOL & 180 & 20 & 2.18   & 1.26  & 114.3  & 11.5   & 1.29 & 0.13 & 113.3&  4.9 & -1.29 &  89.0 \\
2021 Dec 11 21:15--21:43 & FAPOL & 180 & 8  & 2.13   & 1.28  & 91.5   & 17.2   & 0.67 & 0.20 & 86.0 & 11.5 & -0.66 &  84.6 \\
2021 Dec 23 19:27--21:39 & FAPOL & 180 & 28 & 2.06   & 1.34  & 79.5   & 23.2   & 0.70 & 0.11 & 159.8&  7.1 &  0.66 &  -9.7 \\
2022 Jan 24 20:08--20:11 & FAPOL & 180 & 8  & 1.84   & 1.59  & 68.4   & 32.4   & 3.55 & 0.32 & 151.36& 3.6 &  3.55 &  -7.07 \\
2017 Dec 09 12:16--17:39 & PICO &   30  &188 &   1.13 &  0.15 &  201.8 &  19.3  & 0.40 & 0.11 & -4.2 &  7.8 & -0.24 &  64.0 \\
2017 Dec 10 10:58--16:53 & PICO &   30  &144 &   1.11 &  0.14 &  187.9 &  19.2  & 0.17 & 0.11 & 15.4 & 18.3 & -0.16 &  97.5 \\
2017 Dec 11 10:46--16:18 & PICO &   30  &424 &   1.10 &  0.12 &  170.4 &  20.0  & 0.02 & 0.10 & 31.5 & 52.0 &  0.00 & 131.2 \\
2017 Dec 12 12:39--16:32 & PICO &   30  &180 &   1.08 &  0.11 &  149.0 &  22.6  & 0.94 & 0.11 & 67.5 &  7.9 &  0.90 &  8.5 \\
2017 Dec 13 10:24--15:08 & PICO &   30  &172 &   1.07 &  0.09 &  129.7 &  27.1  & 1.92 & 0.10 & 34.5 &  3.1 &  1.89 &  -5.1 \\
     \hline
\multicolumn{14}{l}{$ ^a $ UT at exposure start,$ ^b $ Instrument, $ ^c $Exposure time, $ ^d $ Number of images used to the analysis, $^e$ Median heliocentric distance,}\\
\multicolumn{14}{l}{$ ^f $ Median geocentric distance, $ ^g $ Position angle of the scattering plane, $ ^h $ Median solar phase angle, $^i$ Nightly averaged polarization degree,}\\
\multicolumn{14}{l}{$^j$ Uncertainty of $P$, $ ^k $ Position angle of the strongest electric vector, $^l$ Uncertainty of $\theta_\mathrm{P}$, $^m$ Polarization degree referring to the scattering plan}\\
\multicolumn{14}{l}{$^n$ Position angle referring to the scattering plane. }\\
\multicolumn{14}{l}{We note that the PICO data in this table is the result of reanalysis of data published by \citet{Shinnaka+2018}.}\\
\multicolumn{14}{l}{The web-based JPL Horizon system (\url{http://ssd.jpl.nasa.gov/?horizons}) was used to obtain $r$, $ \Delta$, $ \phi$, and $ \alpha$ in the table.}\\
\end{tabular}
\end{table*}

\section{Derivation of polarimetric parameters at low phase angles}
\label{sec:polpara}

Table \ref{table:obs&result} summarizes the weighted means of nightly data. We computed the polarization degree and the position angle referring to the scattering plane ($P_\mathrm{r}$ and $\theta_\mathrm{r}$). Fig. \ref{fig:phase-plot} indicates the phase angle dependence of $P_\mathrm{r}$. 

In Fig. \ref{fig:phase-plot}, the data taken with different instruments agree well, indicating that the data reduction processes described in Section \ref{sec:observations} seem to work well to eliminate the instrumental effects. Moreover, Phaethon's profile is in good agreement with (155140) 2005 UD (a dynamical association with Phaethon, \citealt{Ohtsuka+2006}), supporting previous results \citep{Ishiguro2022}.

We fit the data of Phaethon at low phase angles ($\alpha < 30$\degr) by using the Lumme--Muinonen function (L/M, \citealt{Lumme+1993}) and linear-exponential function (L/E, \citealt{Muinonen+2009}). We use the same notations as the one used in \citealt{Cellino+2015}. 
The Markov chain Monte Carlo method implemented in PyMC3 \citep{Salvatier+2016} is employed. 
We set boundary conditions of $h\in [0\,\%\,\mathrm{deg}^{-1}, 1\,\%\,\mathrm{deg}^{-1}$], $\alpha_{0} \in [10\degr, 30\degr]$, $c_1 \in [0, 10]$, and $c_2 \in [0, 10]$ for L/M, and $A\in[10, 20]$, $B\in[15, 25]$, and $C\in[0, 1]$ for L/E. The uncertainties of the optimal parameters are derived in the same manner as \citet{Geem2022}. The fitting results and their uncertainties obtained by using L/E are covered by those of L/M.

\section{Results}
As a result of the data fitting, we obtained the minimum polarization degree $ P_\mathrm{min}=-1.3 ^{+ 0.1 }_{- 0.1 }$\,\% at the phase angle $\alpha_\mathrm{min}=	9.0^{+ 0.7 }_{- 0.8 }\degr$, the polarimetric slope $h= 0.22^{+ 0.01 }_{- 0.02 } $\,\%\,$\mathrm{deg}^{-1}$, and the inversion angle $\alpha_0= 19.9^{+ 0.3 }_{- 0.3 }\degr$. As shown below, we further examined the composition and geometric albedo with this result.

\label{sec:results}
\subsection{Comparison with meteorites and other asteroids}

Fig. \ref{fig:h-Pmin} compares $P_\mathrm{min}$ and $h$ of Phaethon with those of carbonaceous chondrites and other C-complex asteroids. As described in \citet{Ishiguro2022}, anhydrous meteoritic samples (CK, CO, and CV) are distributed in the upper left, while hydrous ones (CM and CI) are in the lower right. Because the distribution of Ch-type asteroids (defined by	the	presence of	an	absorption	near 0.7\,$\mu$m due	to	Fe-bearing	phyllosilicates) mostly matches the hydrous meteorite samples, this $P_\mathrm{min}$--$h$ plot is adaptable to actual asteroids. Since the low albedo B-type asteroids (the so-called Themis group, \citealt{Clark+2010}) are distributed between hydrous and anhydrous, their surfaces likely experienced some degree of dehydration. Both Phaethon and 2005 UD are located near the concentration of anhydrous samples and (2) Pallas (B-type with a moderately high albedo, \citealt{Clark+2010}) but significantly deviated from the concentration of hydrous samples. Therefore, we conclude that the surface of Phaethon is likely composed of anhydrous carbonaceous material. Although the anhydrous nature was suggested based on the spectral studies \citep{Clark+2010,Clark+2011,2012Icar..218..196D,Takir+2020}, it is significant that the independent approach via polarimetry ensures the possibility of anhydrous nature.

\subsection{Geometric albedo}
It is known that the geometric albedo in $V$-band ($p_\mathrm{V}$) has a tight correlation with $h$ \citep{Geake+1986}. This relationship is expressed as $\log_{10} \left( p_\mathrm{V} \right) = C_1 \log_{10} \left( h \right) + C_2 $, where $ C_\mathrm{1} $ and $C_\mathrm{2} $ are constants.
These constants are derived using databases of asteroid polarimetry and albedos. We employed the constant values from two recent works  \citep{Cellino+2015,Lupishko+2018}. \citet{Cellino+2015} derived these constants for asteroids without albedo constraint and with $p_\mathrm{V}>0.08$, while \citet{Lupishko+2018} derived these constants without albedo constraint. We used three sets of these constants and estimated the $R_\mathrm{C}$-band geometric albedo of  $p_\mathrm{R_\mathrm{C}}=0.09 \pm 0.01$ for the constants in \citet{Cellino+2015} (without the albedo constraint) and \citet{Lupishko+2018}, and $p_\mathrm{R_\mathrm{C}}=0.11 \pm 0.02$ for the constants in \citet{Cellino+2015} (with the albedo constraint). We regard $p_{R_\mathrm{C}}=p_\mathrm{V}$ because Phaethon has a nearly flat spectrum over this wavelength.
With these $p_\mathrm{V}$ values and errors, the median, minimum, and maximum values are $p_\mathrm{V}=0.11$, 0.08, and 0.13.

\section{Discussion}
\label{sec:discussion}

Phaethon's geometric albedo had been derived by various methods, yet it varies widely from 0.037 to 0.220 in the literature when the errors are considered \citep{Green+1985, Harris+1998, Tedesco+2004, Usui+2011, Hanus+2018, McAdam+2018, Ali-Lagoa+2018,Masiero+2019}. This factor of $\sim 6$ difference made it difficult to establish the fly-by observation plan for the {\it DESTINY$^+$} mission. This large discrepancy may be caused by different thermal models with different absolute magnitudes. Polarimetry has the advantage of converting directly from $h$ to $p_\mathrm{v}$ without any assumptions. It is worth noting that we considered all possible uncertainties (i.e., in $h$, $C_\mathrm{1}$, and $C_\mathrm{2}$) for deriving the reliable $p_\mathrm{v}$ and its range. Although the median albedo value is not so different from previous works, it is significant to narrow the possible range to 1/3 of the previous estimate range for preparing the {\it DESTINY$^+$} fly-by observation. The updated albedo value is also valuable for considering the nature of the asteroid.

The association of Phaethon with comet nuclei has been discussed. From the visible spectrum, Phaethon is classified as either B- (based on \citealt{Bus+2002}) or F-type (based on \citealt{Tholen+1989}). Although F-type asteroids account for only 3\,\% of all asteroids in the Tholen classification, they show an interesting polarization property. \citet{Belskaya2005} noticed that three F-type asteroids exhibited unique $\alpha_0$ values (14--16\degr), which are predominantly smaller than asteroids in general ($\alpha_0\sim20 \degr$). The small $\alpha_{0}$ values of F-types may be linked to two comets, (7968) Elst-Pizarro (i.e., main-belt comet) and 2P/Encke ($\alpha_0$=17.6$\pm$2.1\degr\ in $R$-band and $\sim 13$\degr, respectively, \citealt{Bagnulo2010,Boehnhardt2008}). While the number of comet samples is only two, \citet{Cellino2018} suggested a connection between F-type and comet nuclei. \citet{Belskaya2005} suggested a possible interpretation that an optical homogeneity of regolith microstructure at scales of the order of visible light wavelengths may be responsible for the small $\alpha_0$. 
However, Phaethon's $\alpha_0$ is different from F-type asteroids and two comet nuclei but consistent with asteroids in general. The geometric albedo determined in this study ($p_{V}\sim 0.11$) is significantly higher than comets (including Elst-Pizarro, $p_{V}=0.06-0.07$, \citealt{Boehnhardt2008}) and F-type asteroids ($p_{V}=0.058 \pm  0.011$, \citealt{Belskaya+2017}). Comets generally have a red spectra, while Phaethon has a flat or even blue spectrum. Accordingly, Phaethon's surface materials are likely different from ordinary comets.

How can we explain Phaethon's recent activity \citep{Jewitt+2010} and its anhydrous nature found in our study? From polarimetry at large phase angles, \citet{Ito+2018} suggested that (1) Phaethon's actual albedo could be much lower than the estimate at the time, or (2) the asteroid was covered with large grains (probably produced via a sintering effect at the perihelion). With the updated albedo, we estimated the particle size using the same method as \citet{Ito+2018} and found that it is $\sim$ 300 $\mu$m (with an error of $\sim$ 70 $\mu$m). This particle size is larger than other asteroids such as Ryugu \citep{Kuroda+2021}, increasing the confidence of the sintering hypothesis. 

 Dust ejection under such a high-temperature condition has also been studied recently. \cite{Masiero2021} devised a mechanism for dust ejection by sodium sublimation. \citet{Bach2021} developed an idea of \citet{Jewitt+2010} and proposed that dust production and ejection would happen by the combination of thermal fatigue, thermal radiation pressure from the surface, and solar radiation pressure. These recent studies considered a cometary activity in the high-temperature environment ($\sim 1000$\,K) near the Sun, completely different from general comets, whose activities are driven by ice sublimation. Under such an environment at high temperatures, dehydration \citep{Hiroi+1996} and subsequent sintering would happen near the perihelion. To sum up our findings and other recent research on Phaethon, the surface is unlikely primordial but experiences a high degree of thermal alternation.

\section{summary}
We conducted the polarimetric observations of Phaethon at low phase angles and found that this asteroid has a polarimetric property similar to anhydrous chondrites. Phaethon's albedo and inversion angle are significantly different from comet nuclei. Although the interior composition is still unknown, we conjecture that the surface material shows considerably-evolved features that have experienced thermal metamorphism and dehydration rather than primitive features of comets and hydrous asteroids.

 
\section*{Acknowledgments}
Research activity at Seoul National University was supported by the NRF funded by the Korean Government (MEST) grant No. 2018R1D1A1A09084105. This research was partially supported by the Optical \& Near-Infrared Astronomy Inter-University Cooperation Program, MEXT, of Japan. The observations at NHAO were conducted as an open-use program. SH was supported by the Hypervelocity Impact Facility, ISAS, JAXA. Partly based on observations made with the Nordic Optical Telescope, owned in collaboration by the University of Turku and Aarhus University, and operated jointly by Aarhus University, the University of Turku and the University of Oslo, representing Denmark, Finland and Norway, the University of Iceland and Stockholm University at the Observatorio del Roque de los Muchachos, La Palma, Spain, of the Instituto de Astrofisica de Canarias, and the data was obtained with ALFOSC, which is provided by the Instituto de Astrofisica de Andalucia (IAA) under a joint agreement with the University of Copenhagen and NOT. This research was partially supported by Japan Society for the Promotion of Science (JSPS) KAKENHI Grant-in-Aid for Scientific Research (Early-Career Scientists), 20K14541. We appreciate Dr. Irina Belskaya for providing $ \alpha_0 $--$ P_\mathrm{min} $ values of asteroids. 

 \section*{Data Availability}\label{dataave}
The observational data are available in Zenodo\footnote{\url{https://doi.org/10.5281/zenodo.6791884}}. The source codes and scripts for the data analyses, plots and resultant data tables are available via the GitHub service\footnote{\url{https://github.com/Geemjy/Geem_etal_MNRAS_2022.git}}.

\section*{NOTE ADDED IN PROOF}
We calculated the geometric albedo using recently published results in \citet{Kiselev+2022} with ours and updated it to $p_{V} = 0.10$ (in the range of $0.08$--$0.12$).



\bibliographystyle{mnras}
\bibliography{references.bib} 



\bsp	
\label{lastpage}
\end{document}